\documentclass[preprint,preprintnumbers,amsmath,amssymb]{revtex4}
\usepackage{amsfonts}
\usepackage{amssymb}
\usepackage{latexsym}
\usepackage{graphicx}
\usepackage{psfig}

\newcommand{\al}{\alpha}

\newcommand{\pa}{\partial}
\newcommand{\ep}{\epsilon}
\newcommand{\si}{\sigma}

\newcommand{\la}{\lambda}

\newcommand{\ta}{\tau}
\newcommand{\ga}{\gamma}
\newcommand{\om}{\omega}
\newcommand{\de}{\delta}

\newcommand{\De}{\Delta}

\newcommand{\rar}{\rightarrow}

\def\fun#1#2{\lower3.6pt\vbox{\baselineskip0pt\lineskip.9pt
  \ialign{$\mathsurround=0pt#1\hfil##\hfil$\crcr#2\crcr\sim\crcr}}}
\newskip\humongous \humongous=0pt plus 1000pt minus 1000pt

\newif\ifdtup

\def\cal{\mathcal}
\newcommand{\non}{\nonumber}

\begin{document}

\title{Quartic Anharmonic Many-body Oscillator}

\author{Alexander Turbiner\footnotemark \footnotetext{
\uppercase{O}n leave of absence from the \uppercase{I}nstitute
for \uppercase{T}heoretical and \uppercase{E}xperimental \uppercase{P}hysics, \uppercase{M}oscow 117259, \uppercase{R}ussia.
\uppercase{E}-mail: turbiner@@nuclecu.unam.mx}
}

\address{Instituto de Ciencias Nucleares, UNAM, Apartado Postal 70--543,
\\04510 Mexico D.F., Mexico}

\begin{flushright}
 M\'exico ICN-UNAM 04-05\\
 June, 2004
\end{flushright}

\pacs{no}

\maketitle

\centerline{Abstract}

Two quantum quartic anharmonic many-body oscillators are
introduced. One of them is the celebrated Calogero model (rational
$A_n$ model) modified by quartic anharmonic two-body interactions
which support the same symmetry as the Calogero model. Another
model is the three-body Wolfes model (rational $G_2$ model) with
quartic anharmonic interaction added which has the same symmetry
as the Wolfes model. Both models are studied in the framework of
algebraic perturbation theory and by the variational method.

\vspace{1cm}

\tableofcontents

\newpage

{\sl This work is dedicated to the memory of Ian Kogan who died so
young that is hard for me to imagine that he is not with us
anymore. I knew him for about 30 years since the time when he
appeared at ITEP Theory Division as a young, very brilliant
student. Then for many years we were sitting in the next door
offices at ITEP. Sometimes we talked on science being both
intrigued by the transition from Quantum Mechanics to Quantum
Field Theory. I am sure that Ian would be pleased to read the
present article.}

\section{Introduction}
Anharmonic oscillators play a crucially important role in
contemporary physics since they model intrinsic anharmonic effects
of the real world. The goal of the present work is to introduce and then
to study a special type of quantum anharmonic oscillator -- {\it
many-body anharmonic oscillators}. One of them can be considered
as an anharmonic perturbation of the celebrated many-body Calogero
model (see \cite{Calogero:1971}) or, in other words, the rational
$A_n$ model. Another is an anharmonic perturbation of the three body
Wolfes model or, equivalently, the rational $G_2$ model (see
\cite{Wolfes:1974}). The first system describes $n$ interacting
particles on a line with fixed ordering with pairwise interaction,
while the second one corresponds to three identical interacting
particles on a line with fixed ordering with two- and three-body
interactions. It is rather natural to impose a requirement that
these anharmonic systems should possess the same symmetry
properties as the original Calogero or Wolfes models: (i)
translation invariance, (ii) permutation invariance, (iii)
reflection symmetry with respect to a change of the sign of all
coordinates.

The one-dimensional anharmonic oscillator
\begin{equation}
\label{AH1}
 {\cal H}\ =\ -\frac{d^2}{dx^2} \ +\ x^2 + \la x^4\ ,\qquad x \in R\, ,
\end{equation}
is perhaps one of the most celebrated and the most studied
problems in quantum mechanics. A systematic study was carried out
by Bender-Wu in 1969-1973 in their seminal papers \cite{BW}. Even
this simplest anharmonic oscillator possesses  exceptionally
rich properties:

\begin{itemize}
    \item divergent perturbation theory (PT) in $\la$ \cite{Arkady, BW}, i.e.
      $$E=\sum_k a_k \la^k\ , \qquad a_k \propto k\,!\, ,$$
    \item highly non-trivial but convergent PT in $1/\la$ (strong coupling expansion) \cite{BW, Turbiner:1988},
    i.e. $E=\sum_k b_k \la^{1/3-2k/3}$,
    \item analytic structure in $\la$; all even (odd) eigenvalues are
    analytically related through square-root
    branch points, which  accumulate to $\la=0$ (see Fig.1)  \cite{BW}.

    Hence by
    studying one eigenstate the whole family of eigenstates is
    explored!
\end{itemize}

\vspace{0.1cm}

\vspace{-1.5cm}

\begin{figure}[h]
   \includegraphics*[width=4.8in,angle=0]{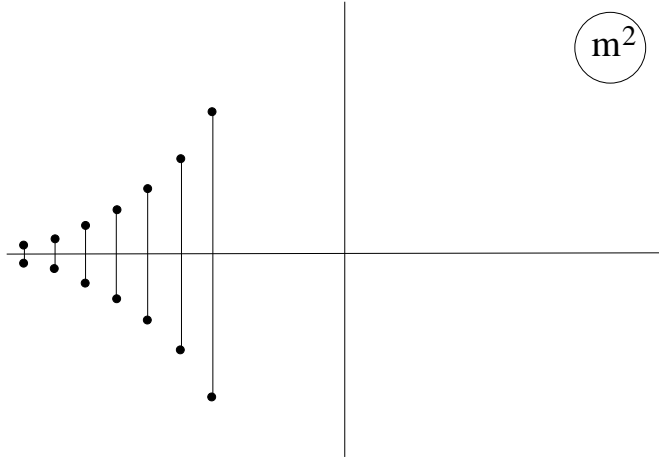}
\caption{Structure of singularities in the inverse coupling
constant $m^2=\la^{-1/3}$ on the first sheet of the Riemann
surface of the ground state energy. Bullets denote the square-root
branch points connected by cuts (vertical lines)}
    \label{fig:1}
\end{figure}

So far very little (almost nothing) is known about eigenfunctions
as functions of $\la$.

\section{Anharmonic Calogero Model}

The Hamiltonian of the $A_{n-1}$ anharmonic oscillator takes the form
\begin{equation}
{\cal H}_{\rm A} = - \frac{1}{2} \sum_{i=1}^{n} \frac{\pa^2}{\pa
{x_i}^2}  + g \sum_{i>j}^{n} \frac{1}{({x_i}-x_{j})^2} +
\frac{\om^2}{2} \sum_{i>j}^{n} (x_i-x_j)^2
 + \frac{\la}{n+6} \sum_{i>j}^{n} (x_i-x_j)^4\, ,
 \label{e1.1}
\end{equation}
where $g > -\frac{1}{4}$, $\om$ is the frequency, $\la \geq
0$ is the coupling constant with a factor $(n+6)$ which is
introduced for a convenience and $n=2,3,\ldots$. This Hamiltonian
describes a system of $n$ identical particles situated on the
straight line with pairwise interaction separated from each other
by impenetrable barriers. The configuration space is
\begin{equation}
\label{e1.2}
 -\infty < x_1 \leq x_2 \leq \ldots \leq x_n < \infty \ .
\end{equation}
If the coupling constant $\la=0$, the Hamiltonian (\ref{e1.1})
corresponds to the celebrated Calogero model and the domain
(\ref{e1.2}) is nothing but the Weyl chamber of the $A_{n-1}$ root
system. The ground state eigenfunction of the Calogero model is
\begin{equation}
\label{e1.2_gs}
 \Psi_{0}^{(c)}(x) = \De^{\nu}(x) e^{-\frac{\om}{2n} X_{2}}\, ,\quad
 g=\nu(\nu-1)\ ,
\end{equation}
where $\De(x) = \prod_{i<j}|x_{i}-x_{j}|$ is the Vandermonde
determinant and $X_{2} = \sum_{i>j}(x_{i}-x_j)^{2}$, when the
ground state energy is $E_{0}^{(c)}=\om(1 + \nu n)$.

In order to deal with translation invariance of many-body systems
we replace the Cartesian coordinates by the center-of-mass
coordinate, $Y=\sum_{j=1}^n x_j$, and the translation-invariant
relative coordinates -- the Perelomov coordinates
\cite{Perelomov:1971},
\begin{equation}
\label{e1.3}
 y_i\ =\ x_i - \frac{1}{n}\, Y\, ,\quad i=1,2,\ldots, n\ ,
\end{equation}
which obey the constraint $\sum_{j=1}^n y_j=0$, where $x_i$ are
the Cartesian coordinates. The coordinates (\ref{e1.3}) make sense
as translation-invariant relative coordinates which measure a
distance from the center of mass to a particle position. Since we consider a
system of identical particles,  permutation symmetry holds. In
order to make manifest the permutation symmetry we introduce permutationally
symmetric coordinates. The most convenient candidate is the
invariants of the symmetric group. Eventually, we arrive at elementary
symmetric polynomials of the arguments $y$ (see Eq.(\ref{e1.3}))
as new coordinates
\begin{equation}
\label{e1.4}
 (x_1,x_2,\ldots x_n) \rightarrow \big( Y, \tau_k (x)=\sigma_k(y(x))|
 \ {\scriptstyle k = 2,3,\ldots n } \, \big)\,.
\end{equation}
Here,
\[
 \si_{k}(x) = \sum_{i_{1}<i_{2}<\cdots<i_{k}}
 x_{i_{1}}x_{i_{2}}\cdots x_{i_{k}}\
\]
are elementary symmetric polynomials. As an illustration let us
present in explicit form the $\ta$ coordinates for $n=2,3$,
putting $-y_n=y_1+y_2+\ldots+y_{n-1}$,
\begin{itemize}
    \item $n=2$ \qquad $\ta_2=-y_1^2$\, ,\\[-2mm]
    \item $n=3$ \qquad $\ta_2=-y_1^2-y_1 y_2-y_2^2\ ,
    \quad \ta_3=-y_1 y_2(y_1+y_2)$\ .
\end{itemize}

It is easy to recognize that the $\ta$ coordinates are nothing but
a particular form of the Weyl invariant polynomials of the lowest
degrees in the $A_{n-1}$ root space.

It can  easily be shown that
\begin{eqnarray}
\label{rel.1}
&& \sum_{i>j}^{n} (x_i-x_j)^2 = -2n \ta_2\, ,\\[1mm]
\label{rel.2}
&& \sum_{i>j}^{n} (x_i-x_j)^4 = 2(n+6) \ta_2^2 - 4n \ta_4\, .
\end{eqnarray}

These relations reveal a remarkable feature of the $\ta$
coordinates -- although the left-hand-side depends on all $x_i$
coordinates, the right-hand-side  depends on a finite number of
$\ta$'s. Making a gauge rotation of the Hamiltonian (\ref{e1.1})
with the Calogero ground state eigenfunction (\ref{e1.2_gs}) as
the gauge factor and re-writing the result in the $\ta$
coordinates (\ref{e1.4}),  we arrive at a strikingly simple
expression after separating out the center-of-mass coordinate $Y$,
\begin{eqnarray}
\label{e1.5}
&& h_{{\rm A}} = 2(\Psi_{0}^{(c)})^{-1}\, ({\cal H}_{\rm A}-E_{0}^{(c)})\,
\Psi_{0}^{(c)} \equiv h_{{\rm Cal}} + \la v_p
\\[1mm]
&&= \sum_{i,j=2}^n {\cal A}_{ij}\, \frac{\pa^2}{\pa {\ta_i} \pa
{\ta_j} } + \sum_{i=2}^n {\cal B}_i \,\frac{\pa}{\pa \ta_i} + 2\la
\Big[ \ta_2^2 - \frac{2n}{n+6} \,\ta_4\Big]\, , \non
\end{eqnarray}
where
\begin{eqnarray}
{\cal A}_{ij}&=&
\frac{(n-i+1)(1-j)}{n}\,\tau_{i-1}\,\tau_{j-1} +\hspace{-10pt}
\sum_{{l\geq}{\max (1,j-i)}} \hspace{-10pt}
(2l-j+i)\,\tau_{i+l-1}\,\tau_{j-l-1} \ , \non
\\[1mm]
{\cal B}_i  &=&
  \bigg(\frac{1}{n}+\nu\bigg){(n-i+2)(n-i+1)}\,\tau_{i-2} +2\om \,i\,
\tau_i \ , \non
\\[1mm]
v_p &=&2\, \ta_2^2 - \frac{4n}{n+6}\,
\ta_4\ .
\end{eqnarray}
Explicit formulae for the first few coefficient functions are
\begin{eqnarray}
&&
{\cal A}_{22}\ = \ 2 \,\ta_2 \, ,\quad {\cal B}_{2}\ =\ 4\,\om\,
\ta_2 + (1+\nu n)(n-1)\,,\non
\\[1mm]
&&
{\cal A}_{23}\ = \ 3 \,\ta_3 \, ,\quad {\cal B}_{3}\ =\ 6\,\om\,
\ta_2\,,\non
\\[1mm]
&& {\cal A}_{24}\ = \ 4\, \ta_4 \, ,\quad {\cal B}_{4}\ =\
8\,\om\, \ta_4 + \frac{1}{n}\,(1+\nu n)(n-2)(n-3) \,\ta_2\,, \non
\\[1mm]
&&
{\cal A}_{33}\ = \ 4 \,\ta_4 - 2 \Big(1-\frac{2}{n}\Big)\,\ta_2^2 \,,\quad
{\cal A}_{34}\ = \ 5\,\ta_5 - 2 \Big(1-\frac{3}{n}\Big)\,\ta_2\ta_3 \, ,
\non\\[1mm]
&&
{\cal A}_{44}\ = \ 6\,\ta_6 + 2\,\ta_2\ta_4 - 3 \Big(1-\frac{3}{n}\Big)\,\ta_3^2
\, .\non
\end{eqnarray}

In \cite{Ruhl:1995} it was demonstrated that at $\la=0$ the
gauge-rotated Calogero Hamiltonian $h_{{\rm Cal}}\ $ (\ref{e1.5})
has infinitely many finite-dimensional invariant subspaces
\begin{equation}
\label{fin-dim}
 {\cal P}_k\ =\ \langle {\ta_{\scriptscriptstyle
2}}^{p_2} {\ta_{\scriptscriptstyle 3}}^{p_3}\ldots
 {\ta_{\scriptscriptstyle  n}}^{p_{n}}
 \vert \ 0 \le \Sigma p_i \le k \rangle\ ,\quad  k=0,1,2,\ldots\, .
\end{equation}
These spaces can be embedded one into another,
\[
{\cal P}_0 \subset  {\cal P}_1 \subset {\cal P}_2 \subset \ldots
 \subset  {\cal P}_k  \subset \ldots \ ,
\]
thus forming an {\em infinite flag (filtration)} ${\cal P}$.
Hence one can say that the operator $h_{{\rm Cal}}$ preserves the
flag ${\cal P}$. Another property of $h_{{\rm Cal}}$ is the existence
of a hidden $gl_{n-1}$ algebra. The Hamiltonian $h_{\rm Cal}$
can be written as a second degree polynomial in generators of the
$gl_{n}$ algebra in the totally symmetric representation
$(k,0,0,\ldots ,0)$,
\begin{eqnarray}
\label{reps}
 {\cal J}_i^{\,-} &=& \frac{\pa}{\pa \ta_i},\qquad \quad
i=2,3\ldots n \ , \non \\[1mm]
 {\cal J}_{ij}^{\,0} &=&
\ta_i \,\frac{\pa}{\pa \ta_j}, \qquad i,j=2,3\ldots n \ ,
 \non \\[1mm]
{\cal J}^{\,0} &=& \sum_{i=2}^{n} \ta_i\,\frac{\pa}{\pa \ta_i}-k\, ,
 \non\\
 {\cal J}_i^{\,+} &=& \ta_i {\cal J}^0 =
x_i\, \Big( \sum_{j=2}^{n} \ta_j\,\frac{\pa}{\pa \ta_j}-k \Big),
\quad i=2,3\ldots n \ ,
\end{eqnarray}
in such a way that the generators ${\cal J}_i^+$ do not appear. It
is worth  mentioning that for integer $k$, $n^2$ the generators
(\ref{reps}) possess a common finite-dimensional invariant
subspace ${\cal P}_k$. This is nothing but a finite-dimensional
irreducible representation space of the algebra $gl_k$ in the
realization (\ref{reps}). Therefore the flag  ${\cal P}$ is made
out of irreducible finite-dimensional representation spaces of the
algebra $gl_{n}$ taken in the realization (\ref{reps}). It is
worth  emphasizing that the perturbation potential is itself an
element of the representation spaces (\ref{fin-dim}),
\begin{equation}
\label{perturbation}
  v_p \in {\cal P}_k\ ,\quad k=2,3,\ldots\,.
\end{equation}

\subsection{Perturbation theory (generalities)}

Now let us consider the spectral problem for the operator $h_{{\rm
A}}$,
\begin{equation}
\label{e1.6}
 h_{{\rm A}} \phi = 2\,\ep\, \phi\ .
\end{equation}
The spectral parameter $\ep$ is related to the  energies $E$ of
(\ref{e1.1}) by
\[
 E\ =\ E_0^{(c)} + \ep\ ,
\]
and $\phi$ is related to the eigenfunction of (\ref{e1.1})
through
\[
 \Psi(x) = \phi(x) \Psi_{0}^{(c)}(x)\ ,
\]
where $\Psi_{0}^{(c)}(x)$ is the ground state eigenfunction of the
Calogero Hamiltonian (\ref{e1.1}) at $\la=0$.
We develop perturbation theory for the equation (\ref{e1.6}) in
powers of $\la$,
\begin{equation}
\label{e1.7}
 \phi = \sum_{k=0}^{\infty} \phi_k \la^k\ ,\quad
 \ep = \sum_{k=0}^{\infty} \ep_k \la^k\ \, ,
\end{equation}
which is in fact the Dalgarno-Lewis form of  perturbation
theory \cite{Dalgarno}. It is easy to derive an equation to find the
$k$th correction
\begin{equation}
\label{e1.8}
 (h_{{\rm Cal}} - 2\ep_0) \phi_k = 2 \sum_{i=1}^{k} \ep_i \,\phi_{k-i} -
v_p \,\phi_{k-1} \ .
\end{equation}

Following the theorem from \cite{Turbiner:1999}, as a consequence of
the property (\ref{perturbation}) the perturbation theory
(\ref{e1.7}) is algebraic -- all perturbation corrections $\phi_k$
are polynomials in $\ta$'s of finite degree. Hence the
construction of perturbation theory is a linear algebraic procedure.

\subsection{Perturbation theory (concrete results)}

\vskip 0.3cm

\subsubsection{Ground state}

The ground state of the gauge-rotated Calogero Hamiltonian
$h_{{\rm Cal}}$ is (see (\ref{e1.5}))
\begin{equation}
\label{gr-state_n}
 \phi_0=1\ ,\quad \ep_0=0\ .
\end{equation}
A simple analysis of equation (\ref{e1.8}) shows that the first
eigenfunction correction $\phi_1 \in {\cal P}_2$, hence it should
be a second degree polynomial in the $\ta$'s. After substitution of such
an Ansatz into (\ref{e1.8}) simple calculations give
\[
 \ep_1 = \frac{n(n-1)(1+\nu n)[6 - \nu(6 - 5 n)]}{4(n+6)\om^2}\ ,
\]
\begin{equation}
\label{1corr_n}
 \phi_1=\frac{n}{2(n+6)\om}\, \ta_4 - \frac{1}{4\om} \,\ta_2^2 +
 \frac{n[6 - \nu(6 - 5 n)]}{4(n+6)\om^2}\,\ta_2\ .
\end{equation}
It is worth  mentioning that the correction $\phi_1$ depends on two
$\ta$ variables only, $\ta_{2,4}$.

A similar analysis of equation (\ref{e1.8}) shows that the second
eigenfunction correction $\phi_2 \in {\cal P}_4$, hence it should
be a fourth degree polynomial in the $\ta$'s. After substitution of such
an Anzatz into (\ref{e1.8}) simple calculations give
 for the second correction
\begin{eqnarray}
\label{2corr_n}
&&\ep_2 = -\frac{n(n-1)(1 + \nu n)}{16(n+6)^2\om^5}
 \big[150n+36 + \nu (5n - 6)(49n + 6)\\
&&\hspace{4cm} ~+ \nu^2 n(101 n^2 -245n + 150)\big]\, ,
\non\\[2mm]
&& (n+6)^2\phi_2\ =
\bigg[
 -\frac{n^2}{4{\om}^3} \,\ta_6 + \frac{n^2}{4{\om}^2} \,\ta_4^2
 - \frac{n(n+6)}{4{\om}^2}\, \ta_4\ta_2^2
\non\\
&&~~~~+\frac{n (7n+8-6\nu n+5 \nu n^2)}{4{\om}^3}\, \ta_4\ta_2
- \frac{n [19n+6 + \nu n(14n -19)]}{8\om^4}\, \ta_4
\non\\
&&
~~~~+ \frac{n(n-3)}{8 \om^3} \,\ta_3^2
 + \frac{(n+6)^2}{16\om^2}\, \ta_2^4
 - \frac{(n+6)[4(5n+3) + 3 \nu n(5n-6)]}{24\om^3}\, \ta_2^3
\non\\
&&~~~~+ \frac{55n^2+120n+36+\nu n (74n^2+5n-114)+\nu^2 n^2
(5n-6)^2}{16\om^4} \,\ta_2^2
\non\\
&& ~~~~- \frac{n[150n+36+\nu(49n+6)(5n-6)+\nu^2
n(101n^2-245n+150)]}{16\om^5} \,\ta_2 \bigg]\,.
\non
\end{eqnarray}
It is worth mentioning that the correction $\phi_2$ depends on
four $\ta$ variables only, $\ta_{2,3,4,6}$.

A similar analysis of equation (\ref{e1.8}) shows that the third
eigenfunction correction $\phi_3 \in {\cal P}_6$, hence it should
be a sixth degree polynomial in the $\ta$'s. After substitution of
such an Anzatz into (\ref{e1.8}) after simple straightforward
calculations we get for the third energy correction
\begin{eqnarray}
\label{3corr_n}
&& \hspace{-0.5cm}\ep_3 = \frac{n(n-1)(1 + \nu n)}{32(n+6)^3\om^8}
 \big[18\,(36+144n+215n^2)
\\
&&\hspace{3.3cm}- 9 \nu\, (72+504n+772n^2-1033n^3)
\non\\
&&
 \hspace{3.3cm}+ n \nu^2 (2592+6948n-16524n^2\!-7529n^3)
\non\\
&&
 \hspace{3.3cm}  - n^2 \nu^3
 (3870-9297n+7529n^2\!-2052n^3)\big]\, . \non
\end{eqnarray}
It is worth  mentioning that the correction $\phi_3$ depends on six
$\ta$ variables only, $\ta_{2,3,4,5,6,8}$.

It can be shown that the $k$th correction to the eigenfunction
$\phi_k \in {\cal P}_{2k}$ for $2k+2 \leq n$; hence it should be a
$2k$th degree polynomial in the $\ta$'s. It takes the form
\begin{equation}
\label{kcorr_phi_n}
 \phi_k\ =\ Pol_{2k} (\ta_2,\ta_3,\ldots \ta_{2k+2})
\end{equation}
and depends on $2k$ of the $\ta$ variables only, $\ta_{2,3,4,\ldots, 2k,
2k+2}$. In general, only when $2k+2 \geq n$ does the $k$th correction
begin to depend on all $n$ of the  $\ta$ variables. Hence the first
corrections (which are important in practice) contain very few
$\ta$'s independently of $n$.

The first three energy corrections $\ep_{1,2,3}$ have a quite
non-trivial property -- they vanish at non-physical values of
$n=0,1, -1/\nu$. It seems quite natural to conjecture that the
correction of arbitrary order will continue to have this property so that
\begin{equation}
\label{kcorr_n}
 \ep_k = -\frac{n(n-1)(1 + \nu
n)}{(n+6)^k\om^{3k-1}}\ {\tilde \ep_k (n,\nu)}\ .
\end{equation}
where ${\tilde \ep_k (n,\nu)}$ is a polynomial in $n,\nu$. Most
likely there exist some physical reasons behind of this property,
but the present author is not aware of them.

It is worth  mentioning that there is no doubt that the present
perturbation theory (\ref{e1.7}) is divergent. The coefficients
$\ep_k$ should grow factorially with $k$. However, it is not clear
how to calculate the index of divergence.

If $g=0$ in (\ref{e1.1}), the singular term in the potential
disappears and the formulae for corrections simplify. This happens when
$\nu=0,1$ (see (\ref{e1.2_gs})) for which
\begin{itemize}
\item
 at $\nu=0$,
\[
\ep_1 = \frac{3n(n-1)}{2(n+6)\om^2}\ ,\\
\]
\[
\ep_2 = -\frac{3 n (n-1)(25n+6)}{8(n+6)^2\om^5}\ ,\\
\]
\[
 \ep_3 = \frac{9 n (n-1)(215 n^2 + 144 n +36)}{16(n+6)^3\om^8}\, ,
\]

\medskip

\item at $\nu=1$,
\[
 \ep_1 = \frac{5n^2(n^2-1)}{4(n+6)\om^2} \ ,\\
\]
\[
 \ep_2 = -\frac{n^2(n^2-1)(101n^2+36)}{16(n+6)^2\om^5}\ ,\\
\]
\[
 \ep_3 = \frac{n^2 (n^2-1)(1026 n^4+1035 n^2+324)}{16(n+6)^3\om^8}\ .
\]
\end{itemize}
It is interesting to point out that at $\nu=1$ the numerators of
$\ep_{1,2,3}$ depend on $n^2$. This could be a general feature
hold for arbitrary correction.

For the two-body case, $n=2$, the problem is reduced to a standard
one-dimensional anharmonic oscillator where $\nu=0$ and $\nu=1$
cases correspond to the ground state and the first excited state,
respectively. Explicitly the corrections are,
\begin{itemize}
\item
 at $\nu=0$,
\[
\ep_1 = \frac{3}{8\om^2}\ ,\quad \ep_2 = -\frac{21}{32\om^5}\ ,\quad \ep_3
= \frac{333}{128\om^8}\ ,
\]

\medskip

\item at $\nu=1$,
\[
 \ep_1 = \frac{15}{8\om^2}\ ,\quad
 \ep_2 = -\frac{165}{32\om^5}\ ,\quad
 \ep_3 = \frac{3915}{64\om^8}\ ,
\]
\end{itemize}
in agreement with the results of the calculation carried out in
\cite{BW} and \cite{Turbiner:1984}.

\subsubsection{First excited state}

The first excited state of the gauge-rotated Calogero Hamiltonian
$h_{{\rm Cal}}$ (see (\ref{e1.5})) is characterized by,
\begin{equation}
\label{exc-state_n}
 \phi_0=\ta_2+\frac{(n-1)(1+\nu n)}{4\om}\ ,\quad  \ep_0=2 \om\ .
\end{equation}
A simple analysis of equation (\ref{e1.8}) shows that the first
eigenfunction correction $\phi_1 \in {\cal P}_3$, hence it should
be a third degree polynomial in $\ta$'s. After substitution of such
an Anzatz into (\ref{e1.8}) simple calculations give
\begin{equation}
\label{exc-state_1cor_n}
 \hspace{-1.7cm}\ep_1 = \frac{n[6+\nu (5n-6)][n+11 + \nu n(n-1)]}{4(n+6)\om^2}\ ,
\end{equation}
\[
 \phi_1=\frac{n}{2(n+6)\om}\,\ta_4\ta_2 +
 \frac{n[n-9+\nu n(n-1)]}{8(n+6)\om^2}\,\ta_4 -\frac{1}{4\om} \,\ta_2^3\\
\]
\[
 -\ \frac{n^2-27n-54 + \nu n(n^2-15n+18)}{16(n+6)\om^2}\,\ta_2^2~~~\\
\]
\[
 +\ \frac{ n[6+\nu (5n-6)][n+11 + \nu n(n-1)]}{16(n+6)\om^3}\,\ta_2\ .
\]
It is worth  mentioning that the correction $\phi_1$ depends on
two $\ta$ variables only, $\ta_{2,4}$ (cf. (\ref{1corr_n})).

A similar analysis of equation (\ref{e1.8}) shows that the second
eigenfunction correction $\phi_2 \in {\cal P}_5$, hence it should
be a fifth degree polynomial in $\ta$'s. After substitution of such
an Anzatz into (\ref{e1.8}) simple calculations give the following
results for the second correction
\[
\hspace{-1cm} \ep_2 = \frac{1}{32(n+6)^2\om^5}\big[-3n^4+342n^3+6471n^2+5574n-360\\
\]
\[
\hspace{3.4cm}-\,2\nu n(3n^4-437n^3-4564n^2+3666n+2916)\\
\]
\[
\hspace{3.4cm}-\, \nu^2 n^2 (3n^4-734n^3-2421n^2+9092n-6324)\\
\]
\[
\hspace{2.2cm} +\, 2\nu^3 n^3(n-1)(101n^2-245n+150)\big]\ .
\]
We will not present the explicit form of $\phi_2$ due to its
complexity. It is worth mentioning that the correction $\phi_2$
depends on four $\ta$ variables only, $\ta_{2,3,4,6}$ as
happens for the ground state (see (\ref{2corr_n})). Neither
$\ep_1$ nor $\ep_2$ vanish simultaneously for some value of $n$.
It can be shown that the $k$th eigenfunction correction has the property
$\phi_k
\in {\cal P}_{2k+1}$.

If $g=0$ in (\ref{e1.1}), the singular term in the potential
disappears and the formulae for corrections simplify. This happens when
$\nu=0,1$ (see (\ref{e1.2_gs})) for which
\begin{itemize}
\item
 at $\nu=0\ , \ \phi_0=\ta_2 + (n-1)/4\om$\ ,
\[
 \ep_1 = \frac{3n(n+11)}{2(n+6)\om^2}\ ,\\[-2mm]
\]

\[
 \ep_2 = \frac{3(n^4-114n^3-2157n^2-1858n+120)}{32(n+6)^2\om^5}\, ,
\]

\bigskip

\item at $\nu=1\ ,\ \phi_0=\ta_2 + (n^2-1)/4\om$\ ,
\[
 \ep_1 = \frac{5n^2(n^2+11)}{4(n+6)\om^2}\ ,\\[-2mm]
\]

\[
 \ep_2 = -\frac{199n^6+36n^5+4082n^4+78n^3+5463n^2-258n-360}
{32(n+6)^2\om^5}\ .
\]
\end{itemize}

For the two-body case, $n=2$ the problem is reduced to a standard
one-dimensional anharmonic oscillator where $\nu=0$ and $\nu=1$
cases correspond to the second and third excited states,
respectively. Explicitly the corrections are,
\begin{itemize}
\item
 at $\nu=0$,
\[
\ep_1 = \frac{39}{8\om^2}\ ,\quad \ep_2 = -\frac{615}{32\om^5}\ ,
\]

\item at $\nu=1$,
\[
 \ep_1 = \frac{75}{8\om^2}\ ,\quad
 \ep_2 = -\frac{1575}{32\om^5}\ ,
\]
\end{itemize}
in agreement with the results of the calculation carried out in
\cite{BW} and \cite{Turbiner:1984}.

\subsection{Correlation functions and perturbation theory}

By purely algebraic means we calculated the first correction to
the ground state energy (\ref{1corr_n}). Making a comparison of
this result with a formula for the first energy correction in the
Rayleigh-Schroedinger perturbation theory we find
\cite{Turbiner:1999} that, in fact, we have calculated the
expectation value
\[
\ep_1 \ =\  \frac{\langle 0 \vert v_p  \vert 0 \rangle}{\langle 0
\vert 0 \rangle} =
  \frac{n(n-1)(1+\nu n)[6 - \nu(6 - 5 n)]}{4(n+6)\om^2}\ .
\]
This expectation value is a rational function of the parameters $\om,
n, \nu$
\[
\frac{\displaystyle \int_{D_c} \sum_{i<j}^n (y_i-y_j)^4
\prod_{i<j}^n|y_{i}-y_{j}|^{2\nu}
e^{-\frac{\om}{2n}\sum{(y_i-y_j)^{2}}} d^{n-1}y }{\displaystyle
\int_{D_c} \prod_{i<j}|y_{i}-y_{j}|^{2\nu}
e^{-\frac{\om}{2n}\sum{(y_i-y_j)^{2}}} d^{n-1}y}\ =\\[-2mm]
\]
\[
  \hspace{-2.8cm} =\ \frac{n(n-1)(1+\nu n)[6 - \nu(6 - 5 n)]}{4(n+6)\om^2}\ ,
\]
where $-y_n=\sum_{i=1}^{n-1} y_i$ and the domain of integration is
the principal Weyl chamber. It is quite amazing that although each
integral is a complicated combination of Euler $\Gamma$-functions,
their ratio reduces to the rational function. This is a general
property which appears in algebraic perturbation theory
\cite{Turbiner:1999}.

\subsection{ Variational study}

We consider a strong coupling limit $\la \rar \infty$ in
(\ref{e1.1}), which is equivalent to putting $\om=g=0$. Following
the recipe for choice of the trial functions (see e.g.
\cite{Turbiner:1984}), the simplest trial function for the ground
state can be written as
\begin{equation}
\label{var-psi}
 \Psi_{trial}\ =\ e^{-\al (-\ta_2) - \frac{2}{3}\beta (-\ta_2)^{3/2} -
 \ga (a^2 + \ta_3^2)^{1/2}-
 \de (\ta_2(\frac{2n}{n+6} \ta_4-\ta_2^2 ))^{1/2} }\ ,
\end{equation}
where $\al, \beta,\ga, \de$ and $a$ are variational parameters.
From dimensional arguments it seems clear that the ground state
energy should be of the form
\begin{equation}
\label{var-ener}
 E_n \ =\ f(n) \la^{\frac{1}{3}}\ .
\end{equation}
For two- and three-body cases the result of calculations is
\begin{equation}
\label{v2}
 f(2) \ =\ 0.53042 \quad (\al = 0.837,\ \beta = 0.837,\ a=\ga=\de=0)\ ,\\[-2mm]
\end{equation}
\begin{equation}
\label{v3}
 f(3) \ =\ 1.17273 \quad (\al = 0.914,\ \beta = 0.845,\ a=\ga=\de=0)\ .
\end{equation}
For the two-body case one can make a comparison with the best
numerical studies, $f(2)_{numerics}=0.530362...$ (see e.g.
\cite{Turbiner:1984}). Hence the simple trial function
reproduces four significant digits in the energy.\\[-0.8cm]

\section{Anharmonic Wolfes Model}

Let us introduce the Hamiltonian which describes a system of three
identical particles with two- and three-body interactions
\[
\hspace{-0.5cm}{\cal H}_{\rm G} =
 \frac{1}{2}\sum_{k=1}^{3}\bigg[-\frac{\pa^{2}}{\pa x_{k}^{2}}
+ \om^2 x_k^2 \bigg]\ + g \sum_{k<l}^{3}\frac{1}{(x_{k} -
x_{l})^2}~~~~~~~\\[-5mm]
\]
\begin{equation}
\label{e2.1} \qquad+\, 3g_1\hspace{-10pt}\sum_{ k<l, \, k,l \neq m}^{3}
 \frac{1}{(x_{k} + x_{l}-2x_{m})^2}\ +
 \frac{\la}{36} \sum_{k<l}^{3}(x_{k} - x_{l})^4 \ ,
\end{equation}
where $\om$ is the parameter, $g=\nu (\nu-1) > -\frac{1}{4}$ and
$g_1=\mu (\mu -1) > -\frac{1}{4}$ are the coupling constants
associated with the two-body and three-body interactions,
respectively, $\la \geq 0$ is an anharmonic coupling constant and
the factor $1/36$ is introduced for convenience. We call this
system the {\it $G_2$ anharmonic oscillator.}

At $\la=0$ the Hamiltonian (\ref{e2.1}) becomes the Hamiltonian of
the rational $G_2$ model which was introduced for the first time
by Wolfes \cite{Wolfes:1974} and later  obtained in the
Hamiltonian Reduction method
\cite{Olshanetsky:1977,Olshanetsky:1983}. Its ground state is
given by
\begin{equation}
\label{e2.2}
 \Psi_{0}^{({\rm r})}(x) = (\De_1^{(r)}(x))^{\nu}
 (\De_2^{(r)}(x))^{\mu} e^{-\frac{1}{2}\om \sum_{k=1}^3 x_k^2} \,
 ,\quad E_0 = \frac{3}{2}\om(1+2\nu+2\mu)\,,
\end{equation}
where $\De_1^{(r)}(x)=\prod_{i<j}^3|x_i-x_j|$ and
$\De_2^{(r)}(x)=\prod^3_{i<j; \ i,j\neq k}|x_i+x_j-2x_k|$.

An interesting observation is that all fourth order
permutationally symmetric and translation invariant polynomials
correspond to two body interactions because\\[-6mm]
\begin{eqnarray}
&&
(x_1+x_2-2x_3)^4+(x_1+x_3-2x_2)^4+(x_2+x_3-2x_1)^4 =\non \\
&&
=9\,[(x_1-x_2)^4+(x_1-x_3)^4+(x_2-x_3)^4] \ ,\non \\
&& \hspace{-0.5cm}{\rm and}\nonumber\\
&&
(x_1-x_2)^2(x_1-x_3)^2+(x_1-x_2)^2(x_2-x_3)^2+(x_1-x_3)^2(x_2-x_3)^2=\non \\
&&=1/2\,[(x_1-x_2)^4+(x_1-x_3)^4+(x_2-x_3)^4] \ .\non
\end{eqnarray}
This leads to the important conclusion that a general fourth
degree polynomial translation invariant potential reduces to two
body interactions. Therefore, the Hamiltonian (\ref{e2.1})
describes the most general permutationally symmetric and
translationally invariant anharmonic oscillator associated with
the $G_2$ rational model with fourth order polynomial
anharmonicity.

Let us make a gauge rotation of the Hamiltonian (\ref{e2.1}) with
the ground state eigenfunction (\ref{e2.2}),
\begin{equation}
\label{e2.3}
 h_{G_2} =  2(\Psi_0^{(r)}(x))^{-1}
 ({\cal H}_{G_2}-E_0) \Psi_0^{(r)}(x) \  .
\end{equation}
The result can be written in terms of two relative coordinates and
the  center-of-mass coordinate $X$.

Now let us take the Perelomov relative coordinates (\ref{e1.3})
and introduce new permutationally symmetric relative coordinates,
\begin{equation}
\label{e2.5}
  \la_1 =  y_1^2+ y_2^2 + y_3^2\ ,\quad
  \la_2 =  y_1^2  y_2^2  y_3^2\ ,
\end{equation}
with the condition $-y_3=y_1+y_2$ (cf. (\ref{e1.4})). Making in
(\ref{e2.3}) a change of variables
$$(x_1,x_2,x_3) \rar (Y, \la_1,\la_2)$$ and separating the
center-of-mass motion (and then omitting it), the remaining part
of the Hamiltonian (\ref{e2.3}) takes the form
\[
 h_{\rm G_2}  = -4\la_1\pa^2_{\la_1\la_1} - 24 \la_2\pa^2_{\la_1\la_2}
 - 18\la_1^2\la_2\pa^2_{\la_2\la_2}\\[-2mm]
\]
\begin{equation}
\label{e2.6}
 +\left\{4\om\la_1-4[1+3(\mu+\nu)]\right\}\pa_{\la_1}
 +\left[ 12\om\la_2-9(1+2\nu)\la_1^2\right]\pa_{\la_2}\
 + \la \la_1^2\ .
\end{equation}
This is the {\it algebraic} form of the $G_2$ anharmonic model
(cf. (\ref{e1.5}) at $n=3$). This Hamiltonian possesses a
remarkable property -- among eigenfunctions there exists a family
which depends on the variable $\la_1$ only (!). The ground state
belongs to this family. In order to find the eigenfunctions
depending on $\la_1$ only it is necessary to solve the spectral
problem for the operator
\begin{equation}
\label{e2.7}
 \tilde h_{G}\ =\ -4\la_1 \pa_{\la_1\la_1}^2
 +\left\{4\om\la_1-4[1+3(\mu+\nu)]\right\}\pa_{\la_1}
 + \la \la_1^2 \ .
\end{equation}
By making a gauge rotation the operator (\ref{e2.7}) can be
reduced to the two-body Hamiltonian
\[
{\cal H} = \underbrace{-\frac{1}{2} \sum_{i=1}^2 \frac{
\partial^2}{\partial x_i^2} + \frac{\om^2}{2} (x_1-x_2)^2 +
\frac{[9(\mu+\nu)^2-1/4]}{(x_1-x_2)^2}}_{{\displaystyle
A_1-}\mbox{rational model}} + \frac{\la}{8} \, (x_1-x_2)^4\,.
\]

Similarly to what was done for the $A_n$ anharmonic many-body
oscillator in Section 1.2, one can develop perturbation theory
in powers of $\la$ for the Hamiltonian (\ref{e2.1}) taken in the
algebraic form (\ref{e2.6}).

\section{Conclusion}

We introduced an anharmonic perturbation of two completely-integrable
and exactly-solvable systems, which are in fact anharmonic
many-body oscillators. It is not clear that these systems remain
integrable or whether the anharmonic terms break this feature. However, the
calculation of perturbation corrections is not influenced by
existence or non-existence of integrability. Perhaps, it is
interesting for the $A_n$-anharmonic oscillator to study the limit $n
\rar \infty$ and a field-theoretic limit. Another interesting
question is about the existence of the quasi-exactly-solvable
anharmonic generalizations of the Calogero and Wolves models other
than those found in \cite{Minzoni:1996}.

\section*{Acknowledgments}

Author thanks J.C.Lopez Vieyra for useful conversations, the
interest to the work and a help with computer calculations. The
work is supported in part by DGAPA grant No. {IN124202} and
CONACyT grant {25427-E}.

\def\href#1#2{#2}

\begingroup\raggedright\endgroup


\begin{thebibliography}{10}

\bibitem{Calogero:1971}
         F.~Calogero,
         {J. Math. Phys.} {\bf 12}, 419 (1971).

\bibitem{Wolfes:1974}
         J.~Wolfes,
         J.~Math. Phys. {\bf 15} 1420 (1974).

\bibitem{BW}
         C.~Bender, T.T.~Wu,
         Phys. Rev. {\bf 184}, 1231 (1969);
         Phys. Rev.  D  {\bf 7} , 1620 (1973).

\bibitem{Arkady}
         A.I.~Vainshtein, 'Decaying systems and divergence of the
         series of perturbation theory',\\
         Preprint Nuclear Physics, Novosibirsk (1964) ,\\
         Published
         In *Minneapolis 2002, Continuous advances in QCD* 617-646
         World Scientific

\bibitem{Turbiner:1988}
         A.V.~Turbiner and A.G.~Ushveridze,
         J.~Math. Phys. {\bf 29} 2053 (1988).

\bibitem{Perelomov:1971}
         A.~M.~Perelomov,
         Teor. Mat. Fiz. {\bf 6}, 364 (1971) [Sov. Phys. -- Theor. and Math.
         Phys. {\bf  6}, 263 (1971)].

\bibitem{Ruhl:1995}
         W.~R{\"u}hl and A.~Turbiner,
          Mod. Phys. Lett. {\bf A10}, 2213 (1995) [hep-th/9506105].

\bibitem{Dalgarno}
       A.~Dalgarno and J.T.~Lewis,
       Proc. Royal Soc. {\bf A 233}, 70  (1955).

\bibitem{Turbiner:1999}
    A.V.~Turbiner, {\em Quantum many-body problems in Fock space:
    algebraic forms, perturbation theory, finite-difference analogs},
    The plenary talk given at International FENOMEC Mini-Workshop on Mathematical
    Physics, Mexico-City, Feb.17-20, 1999; \\
    {Yad. Fiz.}  {\bf 65(6)}, 1168 (2002);
    [Physics of Atomic Nuclei {\bf 65(6)}, 1135 (2002)]
       [hep-th/0108160];\\
    {\em Perturbations of integrable
    systems and Dyson-Mehta integrals},  hep-th/0309109.
    Published in  CRM Proceedings and Lecture Notes, 2004.\\

\bibitem{Turbiner:1984}
    A.V.~Turbiner,
    Usp. Fiz. Nauk  {\bf 144}, 35 (1984)
    [Sov. Phys. Uspekhi {\bf 27}, 668 (1984)].

\bibitem{Olshanetsky:1977}
    M.~A.~Olshanetsky and A.~M.~Perelomov,
    {Lett. Math. Phys.} {\bf 2}, 7 (1977).

\bibitem{Olshanetsky:1983}
    M.~A.~Olshanetsky and A.~M. Perelomov,
    {Phys. Rep.} {\bf 94}, 313 (1983).

\bibitem{Minzoni:1996}
   A.~Minzoni, M.~Rosenbaum and A.~Turbiner,
   Mod. Phys. Lett. {\bf A11}, 1977 (1996) [hep-th/9606092].

\end{thebibliography}
\end{document}